\documentclass[journal=nalefd,manuscript=letter]{achemso}

\usepackage{graphicx}
\usepackage{bm}
\usepackage{amsmath}
\usepackage{amsfonts}
\usepackage{amssymb}
\usepackage{braket}
\usepackage{mathrsfs}
\usepackage{color}
\usepackage[colorlinks,linkcolor={black},citecolor={black},urlcolor={black}]{hyperref}
\usepackage{pdfpages}

\newcommand{\bs}{\boldsymbol}
\newcommand{\mc}{\mathcal}

\author{Huiyuan Zheng}
\affiliation{New Cornerstone Science Laboratory, Department of Physics, University of Hong Kong, Hong Kong, China}
\alsoaffiliation{HKU-UCAS Joint Institute of Theoretical and Computational Physics at Hong Kong, Hong Kong, China}

\author{Dawei Zhai}
\affiliation{New Cornerstone Science Laboratory, Department of Physics, University of Hong Kong, Hong Kong, China}
\alsoaffiliation{HKU-UCAS Joint Institute of Theoretical and Computational Physics at Hong Kong, Hong Kong, China}

\author{Cong Xiao}
\email{congxiao@fudan.edu.cn}
\affiliation{Interdisciplinary Center for Theoretical Physics and Information Sciences (ICTPIS), Fudan University, Shanghai 200433, China}
\alsoaffiliation{HKU-UCAS Joint Institute of Theoretical and Computational Physics at Hong Kong, Hong Kong, China}

\author{Wang Yao}
\email{wangyao@hku.hk}
\affiliation{New Cornerstone Science Laboratory, Department of Physics, University of Hong Kong, Hong Kong, China}
\alsoaffiliation{HKU-UCAS Joint Institute of Theoretical and Computational Physics at Hong Kong, Hong Kong, China}

\title{Layer Coherence Origin of Planar Hall Effect: from Charge to Multipole and Valley}

\keywords{Planar Hall effect, In-plane magnetic moment, Twisted bilayers and trilayers}

\begin{document}

\begin{tocentry}
\centering
\includegraphics[width=0.7\textwidth]{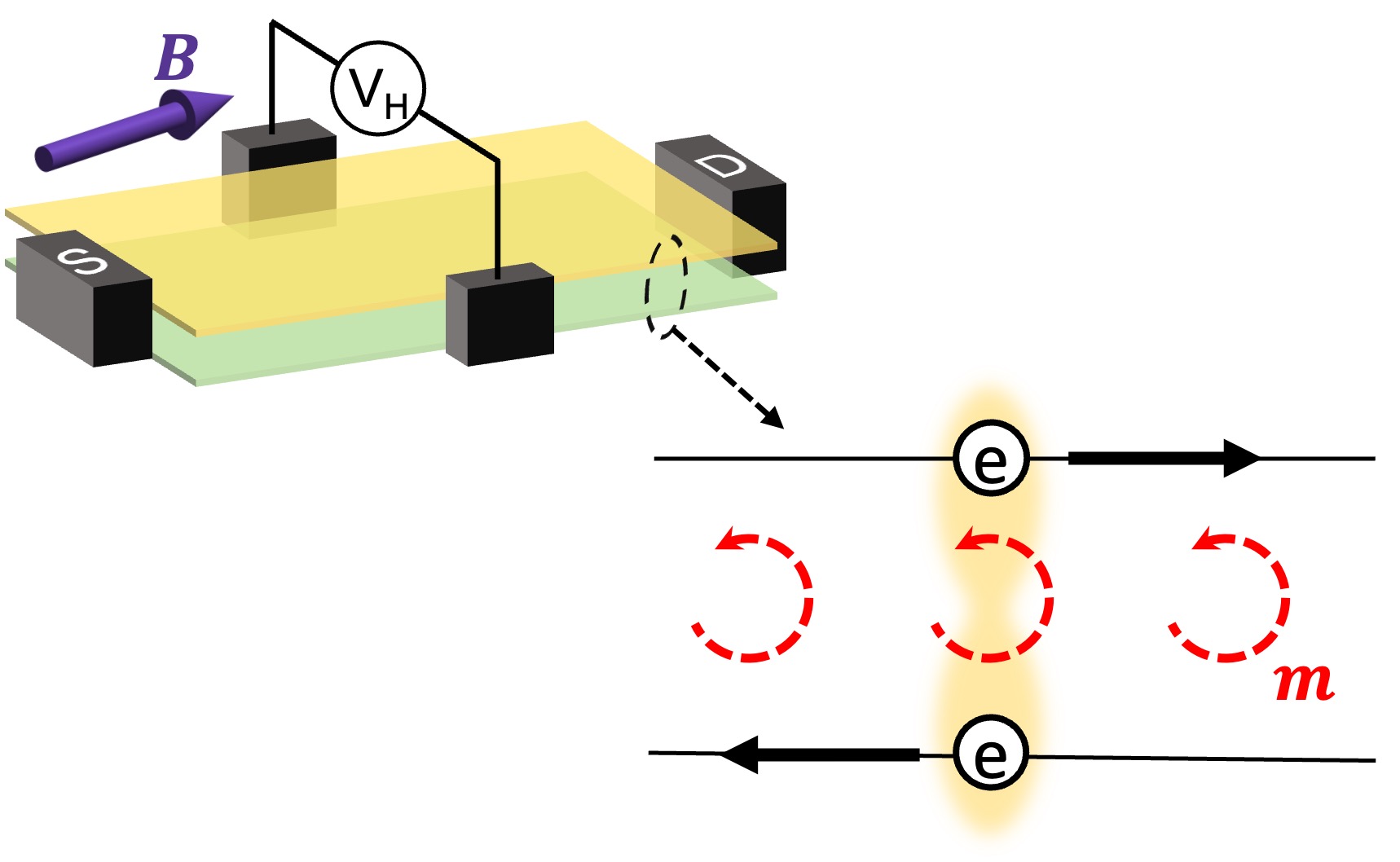}
\end{tocentry}

\begin{abstract}

We uncover a new origin of the planar Hall effect - as an intrinsic property of layer coherent electrons - that exists even in bilayer and trilayer atomically thin limit. It reforms the existing theories requiring three-dimensional orbital motion, or strong spin-orbit coupling of certain forms, which are absent in van der Waals thin films. 
We exemplify that the effect can be triggered by strain and interlayer sliding in twisted structures with rich tunability and strong magnitudes.
Furthermore, this layer coherence mechanism broadens the conceptual framework to include planar multipole Hall effect, and valley Hall effect induced by in-plane pseudo-magnetic field, outreaching the existing mechanisms. The layer mechanism also provides a new route towards quantized Hall response upon a topological phase transition induced by in-plane magnetic field. These results unveil the unexplored potential of quantum layertronics and moir\'e flat band for planar transport in 2D materials.
\end{abstract}


For the exploration of novel transport phenomena, few-layer van der Waals (vdW) materials provide a powerful platform that are responsive to versatile controls by strain, gating and proximity effect.
With the novel role of layer degree of freedom recognized for engineering quantum geometry of electrons~\cite{wu2019topological,yu2019giant,torma2022superconductivity,zhai2023time,ghorai2025planar}, controls unique to the vdW few-layers, including twisting, interlayer sliding and heterostrain, are also added to the toolbox for manipulating new transport functionalities, including various forms of Hall effects~\cite{Xu2021LHE,He2022graphene,Duan2022PRL,Lu2022LHE,Gao2023QM}. In particular, electrons in a most general layer distribution carry interlayer electric multipoles~\cite{zheng2024interlayer},
necessitating the exploration of new forms of current responses beyond the charge component~\cite{fan2024intrinsic}.

\begin{figure}
	\centering
	\includegraphics[width=0.48\textwidth]{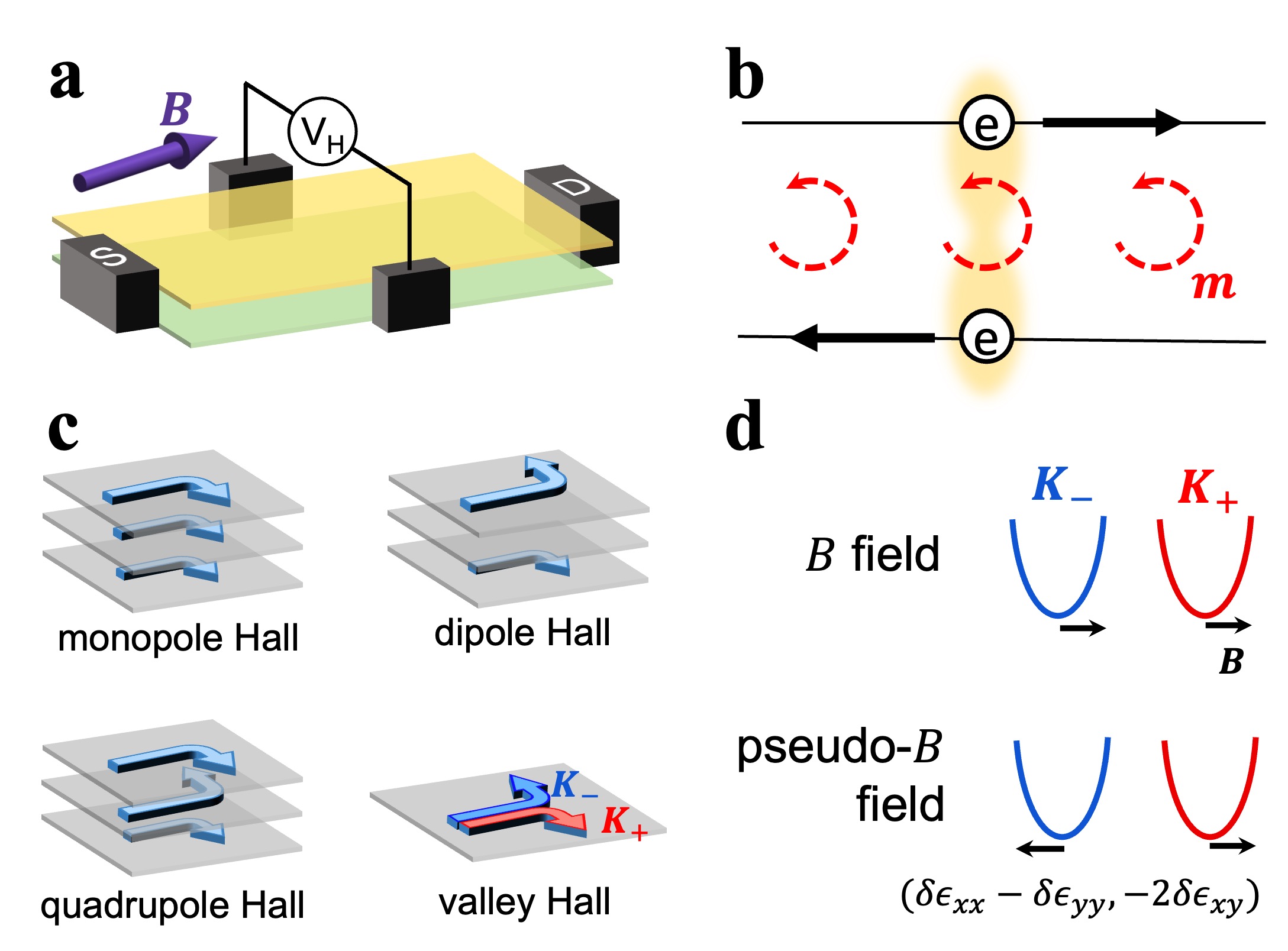}
	\caption{(a) Schematics of planar Hall measurement. (b) In-plane magnetic moment of electron with layer coherent wave function [Eq. (\ref{moment})]. (c) Schematics of planar Hall responses of charge multipoles (the first three). The lower right panel is the valley Hall response. (d) Schematics of vector potential from two valleys. Top: applied magnetic field in-plane. Bottom: valley contrasted in-plane pseudo-magnetic field from heterostrain (see Supporting Information). $\delta \epsilon_{ij}$ is the layer difference of the strain tensor elements.}
	\label{fig_schematic}    
\end{figure}

The intrinsic planar Hall effect (PHE) is a topological transport phenomenon triggered by a magnetic field $B$ applied in-plane (Fig.~\ref{fig_schematic}(a)), initially proposed as a scheme for achieving quantized Hall response \cite{Zhang2011,Liu2013,Ren2016,Liu2018}. 
Recent experimental findings of its un-quantized version supported by the presence of Fermi surface \cite{Ong2018,Zhou2022,Ortix2023} have further stimulated extensive researches on the mechanisms underlying this effect \cite{Zyuzin2020,Ortix2021,Culcer2021,Dai2022,Cao2023,Lu2023,Xiang2024,Su2023,Wang2024IPHE}. 
Two main mechanisms have been proposed based on the magnetic coupling to spin degree of freedom \cite{Zyuzin2020,Ortix2021,Culcer2021,Dai2022,Cao2023,Lu2023} and to the three-dimensional (3D) orbital motion of Bloch electrons \cite{Gao2014,Xiang2024,Su2023,Wang2024IPHE}. 
Intrinsic PHE, however, is apparently not relevant in such vdW platforms. The orbital mechanism known in 3D bulk~\cite{Gao2014,Xiang2024,Su2023,Wang2024IPHE} is quenched in the 2D geometry, while the spin mechanism~\cite{Zyuzin2020,Ortix2021,Culcer2021,Dai2022,Cao2023,Lu2023} is also inactive in the most studied vdW materials including graphene and transition metal dichalcogenides (TMDs)~\cite{moireReviewEvaMacDonaldNatMater2020,moireReviewNatPhysBalents2020,moireReviewRubioNatPhys2021,moireReviewExptFolksNatRevMat2021,moireReviewJeanieLauNature2022,moireexcitonreviewNature2021,moireexcitonreviewNatRevMater2022}. This understanding has impeded the study of intrinsic planar Hall transport in a wide class of materials. 
 Moreover, while studies of Hall effects in out-of-plane fields have extensively extended to the time-reversal symmetric realm addressing quantum degrees of freedom such as spin and valley, the possibility of such paradigmatic shift in the planar geometry has not been envisioned.

In response to these stagnant situations, we uncover a novel intrinsic PHE of layer coherence origin in vdW materials, which works in the few-layer 2D limit without active spin magnetic coupling. This mechanism originates from the lateral motion of out-of-plane charge dipole, which generates an in-plane magnetic dipole that couples to in-plane $B$ field (Fig.~\ref{fig_schematic}(b)). We elucidate the layer-coherence nature of this effect and reveal the underlying band origin, by analyzing the linear-in-$B$ response. This effect is a peculiar property of electrons in layered materials, as an electron wave packet in 3D continuous space carries no charge dipole \cite{Xiao2010,Xiao2021CS}.
In prototypical twisted bilayer and trilayer moir\'e systems, we find sizable PHE over broad parameter spaces, manipulated effectively by heterostrain, interlayer sliding, and interlayer bias.

We further expand the paradigm of PHE in layered structures to charge multipole Hall responses (Fig.~\ref{fig_schematic}(c)), demonstrating the intrinsic dipole Hall effect induced by in-plane $B$ field. Moreover, in time-reversal-preserving case, layer coherence gives rise to intrinsic valley Hall effect, induced by heterostrain that corresponds to in-plane pseudo-$B$
field (Fig.~\ref{fig_schematic}(d)), in centrosymmetric materials that would otherwise forbidden valley transport.
Moreover, the layer origin of in-plane magnetic moment makes possible the first spinless mechanism of quantized PHE, which we demonstrate in twisted trilayer MoTe$_2$.

\emph{Layer-coherence origin of intrinsic PHE.}--In few-layer vdW materials, the interlayer tunneling makes the layer index a quantum pseudospin degree of freedom. Electrons residing in such materials carry an out-of-plane charge dipole $\boldsymbol{\hat{p}}=\hat{p}\boldsymbol{\hat{z}}$. Take a bilayer as example, one has $\boldsymbol{\hat{p}}=-e d_0 \hat{\sigma}_{z}\boldsymbol{\hat{z}}/2$, where $\hat{\sigma}_{z}$ is the Pauli matrix in the layer-pseudospin subspace, and $d_0$ is the interlayer distance. The lateral translational motion of this out-of-plane charge dipole gives an in-plane magnetic dipole, taking an operator form of 
\begin{equation}
    \boldsymbol{\hat{m}}=\frac{1}{2}(\boldsymbol{\hat{p}}\times\boldsymbol{\hat
{v}}-\boldsymbol{\hat{v}}\times\boldsymbol{\hat{p}}),
\label{moment}
\end{equation}
with $\boldsymbol{\hat{v}}$ being the velocity operator. The operator $\hat{\bs{m}}$ can couple to an in-plane magnetic field $\bs{B}$ in the form of $\hat{\mc{H}}_{\mathrm{B}} =-\hat{\bs{m}} \cdot \bs{B}$. Interestingly, $\hat{\mc{H}}_{\mathrm{B}}$ can also be interpreted as a magneto-Stark energy $-\frac{1}{2}(\hat{\bs{p}} \cdot \hat{\bs{E}}_{\text{eff}}+\hat{\bs{E}}_{\text{eff}} \cdot \hat{\bs{p}}) $~\cite{ZhengHuiyuan2022,thomas1961magneto,hopfield1961fine}, where $\hat{\bs{E}}_{\text{eff}} = \hat{\bs{v}} \times \bs{B}$ is the effective out-of-plane electric field felt by electrons in the moving reference frame.
Equation (\ref{moment}) is applicable to the case of $N>2$ as well ($N$ is the number of layers), where the out-of-plane charge dipole is represented by z-component of higher layer pseudospin $\hat{p}=-e d_0\hat{J}^{(\frac{N-1}{2})}_z$. Here the superscript of $\Hat{J}_z$ signifies the pseudospin-$\frac{N-1}{2}$ matrix. For example, in a trilayer, $\hat{p}= -ed_0\text{diag}(1,0,-1)$. The derivation for general N-layers is given in the Supporting Information.

This form of the in-plane magnetic moment operator [Eq. (\ref{moment})] can also be obtained by rigorous quantum mechanical treatment of magnetic field effect, supplied in the Sec.~II of Supporting Information.
With the established in-plane magnetic coupling, an intrinsic PHE is expected in layered vdW systems as thin as the bilayer limit. As $B$ field breaks time-reversal ($\mathcal{T}$) symmetry, a Hall current quantified by the conductivity \cite{Xiao2010} 
$    \sigma_{\mathrm{H}} = -(e^2/\hbar)\sum_n \int[d\bs{k}]f(\tilde{\varepsilon}_n) \tilde{\Omega}_n(\bs{k})
$ can appear,
where $[d\bs{k}]$ is shorthand for $d\bs{k}/ (2\pi)^2$, $n$ is the band index, $f$ is the Fermi-Dirac distribution function, and $\tilde{\varepsilon}$ and $\tilde{\Omega}$ with tilde denote the band energy and $k$-space Berry curvature including the effects of $B$ field. 

To see the intrinsic band origin of the layer induced PHE, we inspect the linear-in-$B$ planar Hall conductivity. Expanding $\tilde{\varepsilon}$ and $\tilde{\Omega}$ to the first order of $B$, the Hall conductivity can be expanded as $\sigma_{\mathrm{H}}=\sigma_{\mathrm{H}}^{(0)}+\sigma_{\mathrm{H}}^{(1)}$ (detailed derivation in Supporting Information). The zero-field value $\sigma_{\mathrm{H}}^{(0)}$ is forbidden by $\mathcal{T}$, whereas the linear-in-$B$ Hall conductivity, which is the dominating contribution, reads
\begin{equation}
\sigma_{\mathrm{H}}^{(1)} = \frac{e^2}{\hbar} \sum_n \int [d\bs{k}]  f'_n \gamma_n(\bs{k}), \label{eq_sigmaH1}
\end{equation}
with 
$
    \gamma_n(\bs{k}) =[\hbar \bs{v}_n \times \bs{\mc{A}}_n^{(1)} - \varepsilon_n^{(1)} \bs{\Omega}_n]_z
$.
Here, $\varepsilon_{n}^{(1)} = -\bs{m}_{n} \cdot \bs{B}$ is the magnetic dipole energy on a Bloch state $\ket{u_{n\bs{k}}}$, and
\begin{equation}
	\bs{\mc{A}}_{n}^{(1)} = - 2\hbar\ \text{Im} \sum_{\ell \neq n} \frac{\bs{v}_{n\ell}\bs{m}_{\ell n}} {(\varepsilon_{n} - \varepsilon_{\ell})^2}\cdot \bm B
	\label{eq:1stOrderCorrection}
\end{equation}
is the $B$-field induced Berry connection via the perturbation of Bloch state,
with $\bs{v}_{n\ell}$ and $\bs{m}_{\ell n}$ being the interband matrix elements of corresponding operators. The tensor
$\alpha_{ab}=\partial \mc{A}_a^{(1)} / \partial B_b$ 
represents an in-plane Berry connection susceptibility, with $a$, $b$ $\in \{x, y\}$. 

An important observation here is that without interlayer coherence, the layer pseudospin and hence the out-of-plane charge dipole $\Hat{p}$ would become a conserved quantity ($p$), and all $\ket{u_{n\bs{k}}}$ are layer eigenstates. If this happens, the in-plane magnetic dipole operator reduces to $ \boldsymbol{\hat{m}}=p\boldsymbol{\hat{z}}\times\boldsymbol{\hat
{v}}$ hence the Berry connection susceptibility becomes the $k$-space Berry curvature $\alpha_{ab}=-\frac{p}{\hbar}\Omega_z\delta_{ab}$. The induced Berry connection is thus along the $B$ field, taking the form of $\bs{\mc{A}}_{n}^{(1)} = -\frac{p}{\hbar}\Omega_z\bm B$. As such,
the two contributions in $\gamma_n(\bs{k}) $ cancel each other (detailed derivation in Supporting Information). One therefore sees that this intrinsic PHE is a unique property of layer coherent (hybridized) electron wave functions. In other words, the layer coherence renders a novel origin of intrinsic PHE.




\begin{figure}
	\centering
	\includegraphics[width=0.48\textwidth]{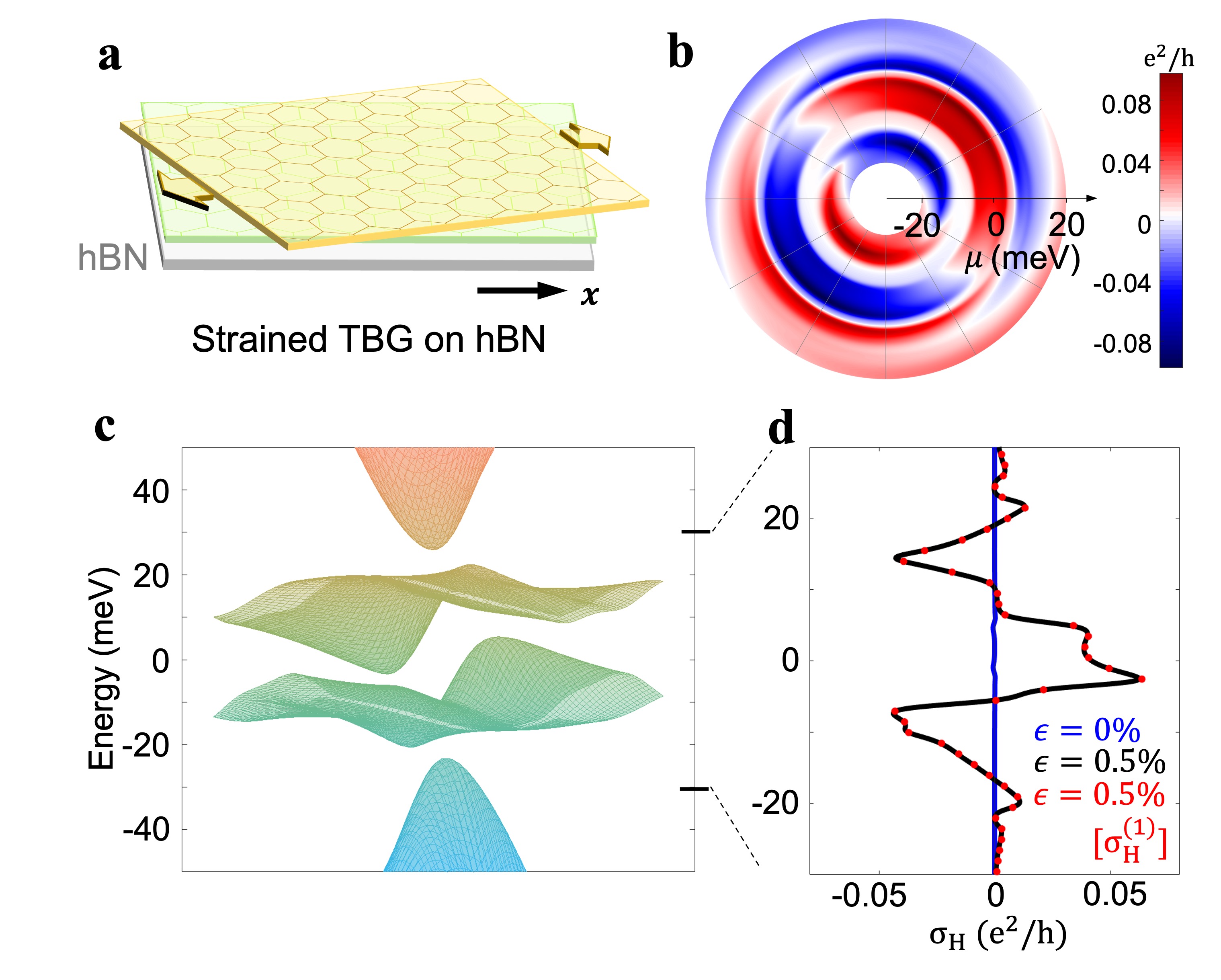}
	\caption{Intrinsic PHE of layer origin in strained TBG on hBN. (a) Schematics of the system. (b) Angular dependence of the intrinsic planar Hall conductivity on $B$ field. The radial direction denotes the chemical potential. (c) Dispersion of $1.2^\circ$ TBG on top of aligned hBN, with $0.5\%$ uniaxial strain applied on the top layer along the $x$ axis. (d) Intrinsic Hall conductivity versus chemical potential, with $B=5$ T applied parallel to the strain. The black (blue) curve is for the case with (without) strain. Red dots describe the $B$-linear response from  Eq. (\ref{eq_sigmaH1}). In the calculation, the temperature is set to 4 K.}
	\label{fig_tbg}    
\end{figure}

\emph{Intrinsic PHE in hetero-strained TBG.}--Symmetry analysis shows that the largest point group that can support the intrinsic PHE is $C_{2h}$ with an in-plane twofold rotation axis. Such a low-symmetry case is relevant in realistic few-layer materials with the presence of strain~\cite{bi2019designing}. 
Here, we use twisted bilayer graphene (TBG) placed on an aligned hBN (Fig.~\ref{fig_tbg}(a)), which is of practical experimental relevance~\cite{sharpe2019emergent,serlin2020intrinsic,gao2021heterostrain} and does not support PHE of spin-orbit origin, as an example to show the layer intrinsic PHE being strong and highly tunable.

The electronic properties in the long-wavelength moir\'e superlattice are described by a continuum model~\cite{koshino2018maximally}, and we consider a hetero-strain applied on the top layer (model details in Supporting Information).
Figure~\ref{fig_tbg}(c) shows the low-energy miniband dispersion around $\bs{K}_+$ valley of $1.2^\circ$ TBG with $0.5\%$ uniaxial strain along $x$ axis, which is defined as the zigzag axis of untwisted monolayer. 
The intrinsic Hall conductivity $\sigma_{\mathrm{H}}$ versus chemical potential in the presence (absence) of strain is shown by the black (blue) curve in Fig.~\ref{fig_tbg}(d), with $B=5$ T applied along $x$ axis. Without strain, the system has $C_{3z}$ symmetry that forbids the effect, which is consistent with the negligible  $\sigma_{\mathrm{H}}$. Red dots in Fig.~\ref{fig_tbg}(d) denote planar Hall conductivity $\sigma^{(1)}_{\mathrm{H}}$. Its excellent coincidence with the black curve indicates the dominance of linear-in-$B$ contribution in the Hall response.

Because Hall conductivity is isotropic with respect to the direction of driving electric field, the linearity in $B$ assures the angular dependence as $\sigma_{\mathrm{H}}=\sigma_{\mathrm{H},x}\cos\varphi+\sigma_{\mathrm{H},y}\sin\varphi$, where $\varphi$ is the angle the $B$ field makes with $x$ axis, and $\sigma_{\mathrm{H},x}$ and $\sigma_{\mathrm{H},y}$ are the Hall conductivity measured when the $B$ field is along the $x$ and $y$ direction, respectively. The angular dependence of $\sigma_{\mathrm{H}}$ (Fig.~\ref{fig_tbg}b) shows that for a wide range of $\varphi$ ($20^\circ \sim 80^\circ$, and $200^\circ \sim 260^\circ$), $\sigma_{\mathrm{H}}$ can reach 0.1$e^2/h$ in the low-energy region around charge neutrality, which is feasible to detect in experiment.
The planar Hall coefficient $\sigma_{\mathrm{H}}/B$ can thus readily reach 22 $(\Omega$cmT)$^{-1}$. This value is nearly two orders of magnitude larger than the predicted intrinsic PHE in semiconductor quantum wells by spin-orbit mechanism \cite{Culcer2021}, and is one order of magnitude greater than the 3D orbital PHE in topological semimetals SrAs$_3$ \cite{Wang2024IPHE} and Fe$_3$Sn$_2$ \cite{Wang2024Fe3Sn2}. Such a large response in TBG emerges from moir\'e flat bands near the small-gap region.

Another advantage of this layer PHE over the effect in 3D materials is its tunability by hetero-strain and by gate, as is detailed in Supporting Information. In particular, for almost any direction of strain, the planar Hall coefficient maintains the large value of 10 $\sim$ 25 $(\Omega$cmT)$^{-1}$ in low-energy regions. Besides, both the sign and magnitude of the effect can be tuned by gate. In the present manuscript, we focus on revealing the previously unexplored layer mechanism for the intrinsic PHE in 2D limit, but how the Hall response evolve as the number of layers is an interesting open question.

\begin{figure}[pb]
	\centering
	\includegraphics[width=0.47\textwidth]{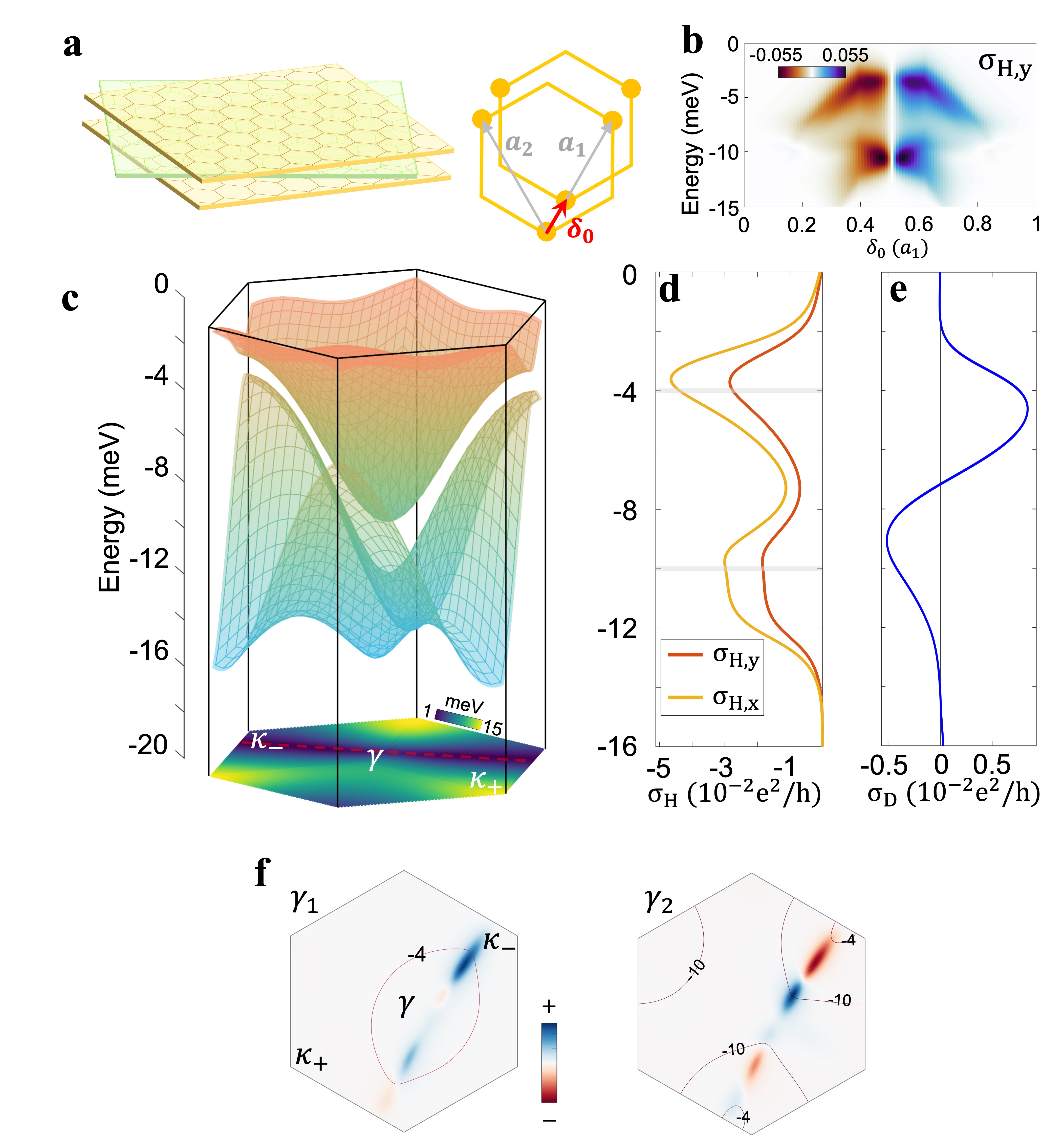}
	\caption{$B$-linear intrinsic PHE in twisted trilayer MoTe$_2$ with interlayer sliding. (a) Schematics of the system, and the top view of the top and bottom layers. (b) Planar Hall conductivity versus the lateral translation, in units of $e^2/h$. The system with translation $\bs{\delta}_0$ is horizontal-mirror symmetric to the one with $\bs{a}_1-\bs{\delta}_0$, so the Hall conductivity is anti-symmetric with respect to the $\bs{\delta}_0 = 0.5 \bs{a}_1$ axis. (c) Low-energy dispersion with twist angle $\theta=3^\circ$ and lateral translation $\bs{\delta}_0 = 0.4\bs{a}_1$. The distribution of energy difference between the two bands is plotted at the bottom, showing a line shaped small-gap region (red dashed line). (d) Planar Hall conductivity as a function of chemical potential. Two curves express the results with $B=5$ T applied in the zigzag ($x$) and armchair ($y$) directions of untwisted layer.  (e) Intrinsic dipole PHE versus chemical potential. (f) Brillouin zone distribution of $\gamma_n (\bs{k})$ [Eq. (\ref{eq_sigmaH1})] of the two bands. Energy contours indicate the Fermi surface at peaks of the response curves in (d).}
	\label{fig_mote2}    
\end{figure}

\emph{Intrinsic PHE in twisted trilayers}.--In trilayers, twisting plus interlayer sliding ~\cite{park2021tunable,PhilipKimTrilayer,tong2020interferences} render a new pathway to tune hot spots of Berry quantities for enlarged layer PHE.
We consider twisted trilayer MoTe$_2$ as an example~\cite{zheng2024interlayer}, which does not allow PHE of spin-orbit origin. Assume the simple case where top and bottom layers are aligned in the same orientation while the middle layer is twisted by a small angle $\theta$ (see Supporting Information) (Fig.~\ref{fig_mote2}(a)). When the two outer layers are fully aligned, the system possesses a $C_{3z}$ axis and a horizontal mirror plane. By introducing a lateral translation between the top and bottom layers ($\bs{\delta}_0$), both symmetries are broken, enabling the intrinsic PHE to appear. 

In Fig.~\ref{fig_mote2}(c), we plot the band dispersion around $\bs{K}_+$ valley in $3^\circ$ twisted trilayer MoTe$_2$ with $\bs{\delta}_0 = 0.4\bs{a}_1$. The first two valence bands are separated by small gaps along a straight line in $k$ space, which can be seen in the energy difference distribution at the bottom of Fig.~\ref{fig_mote2}(c). 
The layer induced planar Hall conductivity as a function of chemical potential is presented in Fig.~\ref{fig_mote2}(d). The two peaks are related to the enhanced Berry connection susceptibility and Berry curvature, which are amplified by the line shaped small-gap region between the first two valence bands, as indicated by the $k$-space distribution of $\gamma_n (\bs{k})$ in Fig. \ref{fig_mote2}(f). At a low hole-doping level around 4 meV, the effect can reach 0.05 $e^2/h$, which is sizable and comparable to that in strained TBG.


The variation of intrinsic PHE with lateral translation is shown in Fig.~\ref{fig_mote2}(b). 
With $\bs{\delta}_0$ growing from $0$, the Hall conductivity gets larger as the breaking of $C_{3z}$ becomes more pronounced. Noticeably, the effect exhibits a suppression at $\bs{\delta}_0 = 0.5\bs{a}_1$ as well as strong enhancement just around it. This is because the gap between the first two valence bands closes at $\bs{\delta}_0 = 0.5\bs{a}_1$ and is slightly opened for translation values around $0.5\bs{a}_1$, which serves as hot spots of Berry quantities. Figure~\ref{fig_mote2}(b) shows that the planar Hall conductivity exhibits a detectable value across a broad sliding range (0.2–0.8 of the moir\'e periodicity). Therefore, in practice, despite lacking a precise control of the lateral sliding, PHE can be observed over a wide range of lateral displacements.

\emph{From charge to multipole and valley current}.-- The layer mechanism not only uncovers the first avenue for intrinsic PHE in widely studied 2D materials with simple spin textures, but also opens up other possibilities in 2D transport studies. 

The hidden layer-resolved pattern of planar transport is a unique feature of layer mechanism, and points
to the rich prospect of layertronic devices~\cite{Jiang2024dissipationless}.
In a vdW layered material, one can naturally ask the question of how the charge current is distributed among the layers, and a general layer distribution can always be expanded as monopole + dipole + quadrupole and so forth.
A complete description of planar Hall conductivity in the layered context hence comes in terms of monopole (charge) Hall part $\sigma_H$, and the multipole (charge neutral) Hall part (Fig.~\ref{fig_schematic}(c)). We exemplify the planar multipole Hall to the order of dipole, and connect it to the in-plane Berry connection susceptibility $\alpha_{ab}$ in Sec. VI of Supporting Information.

To completely describe the planar Hall transport, the dipole Hall current is calculated in twisted trilayer MoTe$_2$ (Fig.~\ref{fig_mote2}(e)), which represents the difference between the currents of the two outer layers (Fig.~\ref{fig_schematic}(c)).
Remarkably, we prove that this dipole PHE is also a characteristic exclusively possessed by layer-coherent electron wave functions (see Supporting Information).
This effect is the counterpart in nonmagnetic layers of the intensively studied intrinsic layer Hall effect triggered by magnetic order in antiferromagnetic vdW layers~\cite{Xu2021LHE,Lu2022LHE,fan2024intrinsic}.

Moreover, the planar Hall effect by layer coherence can have a time-reversal symmetric counterpart, with charge Hall current replaced by pure valley Hall current. Consider a heterostrain which can also be effectively described as a layer dependent vector potential $\bs{A}_{S,l} = (\epsilon_{l,xx}-\epsilon_{l,yy},-2\epsilon_{l,xy})$~\cite{bi2019designing}(Fig.~\ref{fig_schematic}(d)). This has essentially the same effect as a magnetic field applied in-plane, except that time-reversal symmetry dictates the pseudo-field to have opposite sign at the two valleys, leading to valley contrasted Hall current. We formulate the resulting valley Hall effect in Sec. VII of Supporting Information and evaluate it in a centrosymmetric 2D material, which intrinsically forbids the valley Hall effect (c.f. Fig. S3). This effect extends valleytronics to inversion symmetric materials. It also renders a new type of nonlinear valley Hall response \cite{Yu2014NVHE,2024NVHE} in the bilinear order of electric and mechanical fields, adding a member to crossed nonlinear transport unique to 2D materials \cite{chen2024crossed}.

\begin{figure}[b]
	\centering
	\includegraphics[width=0.48\textwidth]{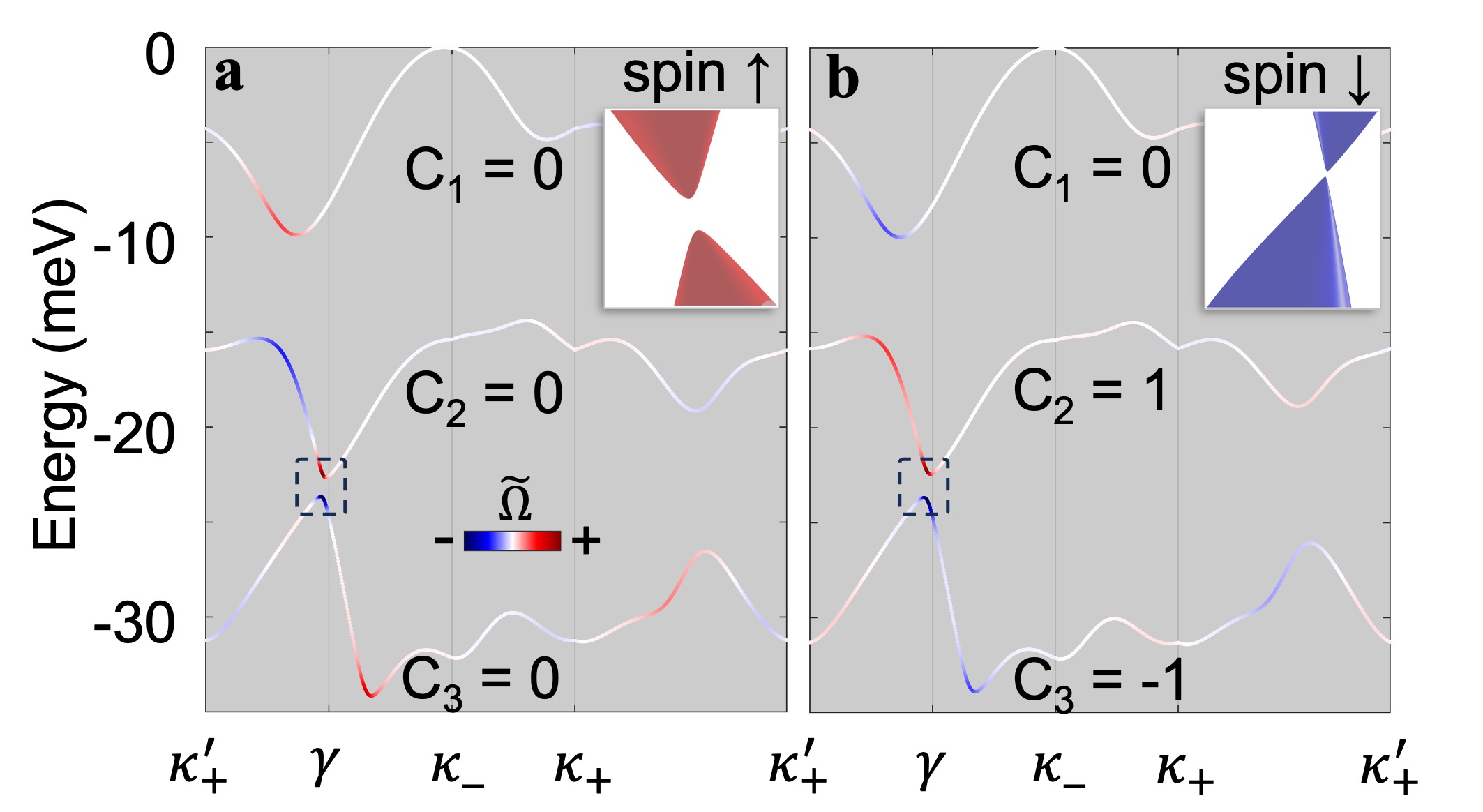}
	\caption{Topological phase transition in twisted trilayer MoTe$_2$. Here an interlayer bias of -9.5 meV is applied to the setup in Fig.~\ref{fig_mote2} with $B$ field of 10 T along $\varphi=120^\circ$. (a,b) Energy dispersion of spin up (a) and down (b) with Berry curvature as the color. The insets are 3D energy dispersion in the dashed boxes in (a) and (b), whose colors carry no information.}
	\label{fig_QAH}    
\end{figure}

\emph{Towards quantizated PHE}.--The layer mechanism can also allow the in-plane $B$ field to open a topological gap, thereby introducing a quantized PHE in the spinless limit, which provides a new route distinctive from the conventional spin-orbit approach \cite{Zhang2011,Liu2013,Ren2016,Liu2018}.

We demonstrate this possibility with the twisted trilayer MoTe$_2$ studied in Fig.~\ref{fig_mote2}. With an interlayer bias of -9.5 meV applied, the local gap between the second and third bands near $\gamma$ point has a small value of 0.1 meV. With a $B$ field of 10 T applied in-plane along  $\varphi=120^\circ$, a topological band inversion occurs for spin-up bands, accompanied by a sign reversal of the Berry curvature in proximity (dashed box in Fig.~\ref{fig_QAH}(a)). As a result, the Chern numbers of the second and third spin-up bands change from $\{-1,1\}$ to $\{0,0\}$, while those for spin down remain $\{1,-1\}$ (Fig.~\ref{fig_QAH}(a) v.s. \ref{fig_QAH}(b)).
The small gap possesses practical challenges to observe the quantized Hall conductivity, which requires low temperature as well as alignment between the gaps of two spin species.
In reality, small out-of-plane tilting of magnetic field~\cite{Ong2018,Zhou2022,Ortix2023} can be exploited to tune the alignment between the gaps of two spin species through Zeeman effect, since these layered vdW materials possess strong g-factors in the out-of-plane direction. On the other hand, at higher temperature and in the absence of a global gap, the topological band inversion by the in-plane magnetic field can manifest as a pronounced peak in the PHE (c.f. Fig.~S2 in Supporting Information).

\begin{figure}[t]
	\centering
	\includegraphics[width=0.48\textwidth]{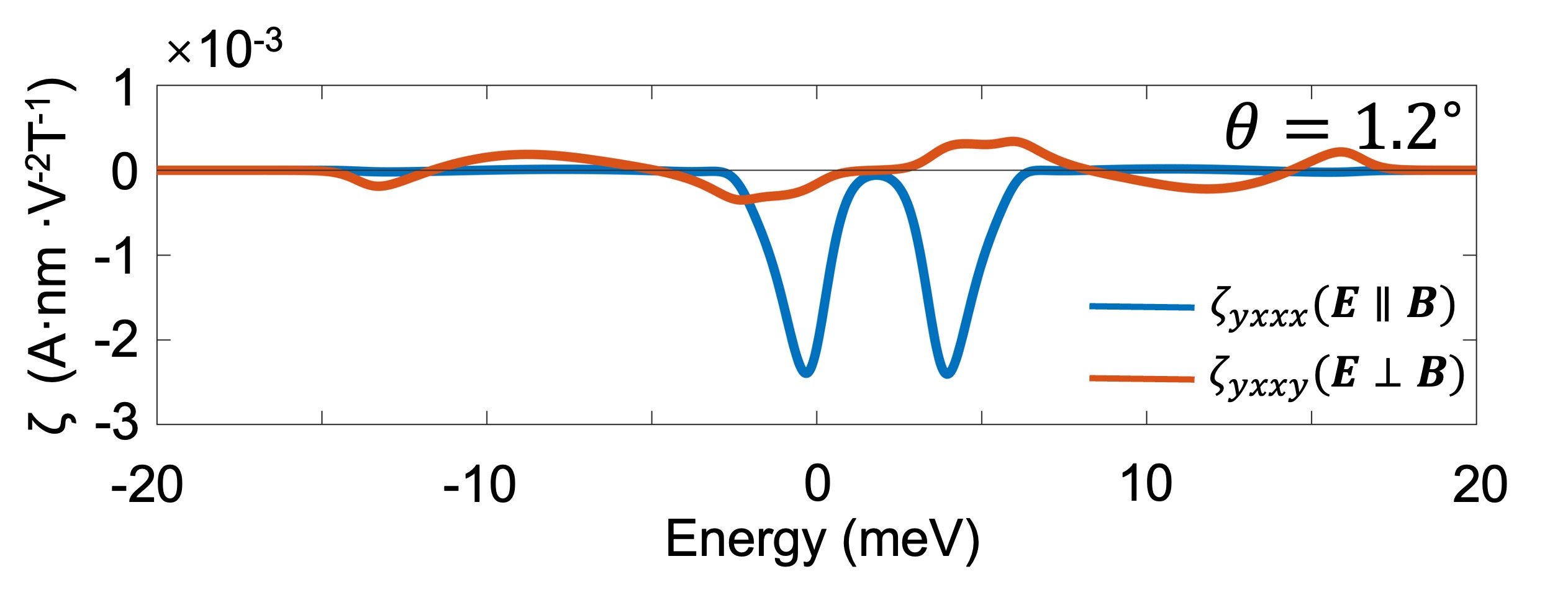}
	\caption{Layer induced intrinsic nonlinear planar Hall coefficient for $1.2^\circ$ TBG on hBN.}
	\label{fig_EEB}    
\end{figure}

\emph{Nonlinear PHE of layer origin}.--The intrinsic PHE has recently been extended to the second order of driving electric field, i.e., the intrinsic nonlinear PHE~\cite{huang2022intrinsic,Ye2023Te}. Here, from the semiclassical theory, we obtain the intrinsic current response of layer origin in the $E^2B$ order (see derivation in Supporting Information): $j^{(2)}_a = \zeta_{abcd} E_b E_c B_d$, where $a,b,c,d$ $\in \{x, y\}$. Symmetry analysis shows that, in contrast to the linear-in-$E$ effect, the nonlinear PHE is allowed in the presence of $C_{3z}$ symmetry, and is quantified by
$
\zeta_\text{H} = \zeta_{yxxx} \text{cos} \psi + \zeta_{yxxy} \text{sin} \psi,
 $
where $\psi$ is the relative angle between the $E$ and $B$ fields. From the calculation based on continuum model of TBG on aligned hBN, we find (Fig.~\ref{fig_EEB}) when $E\parallel B$, $\zeta_\text{H}$ around the charge neutrality is an order of magnitude larger than the effect of spin origin in Janus monolayer MoSSe \cite{huang2022intrinsic}. This further points to rich possibilities to explore other forms of nonlinear transport by the layer mechanism in planar magnetic field in future studies.


\begin{acknowledgement}
This work is supported by the National Key R\&D Program of China (2020YFA0309600), and Research Grant Council of Hong Kong SAR (AoE/P-701/20,HKU SRFS2122-7S05), and the New Cornerstone Foundation.
\end{acknowledgement}


\begin{suppinfo}
The Supporting Information is available free of charge at http://pubs.acs.org.

I. Out-of-plane charge dipole and in-plane magnetic dipole in N-layers; II. Derivation of B-linear intrinsic PHE of layer origin; III. Role of interlayer coherence in intrinsic PHE; IV. More details of the intrinsic PHE in TBG and twisted trilayer MoTe$_2$; V. Moir\'e primitive lattice vectors in real and reciprocal spaces; VI. Intrinsic dipole Hall effect; VII. Heterostrain-induced intrinsic valley Hall effect; VIII. Quantized PHE in twisted trilayer MoTe$_2$; IX. Formalism of layer induced intrinsic nonlinear PHE. 
\end{suppinfo}

\bibliography{reference}

\includepdf[pages=-]{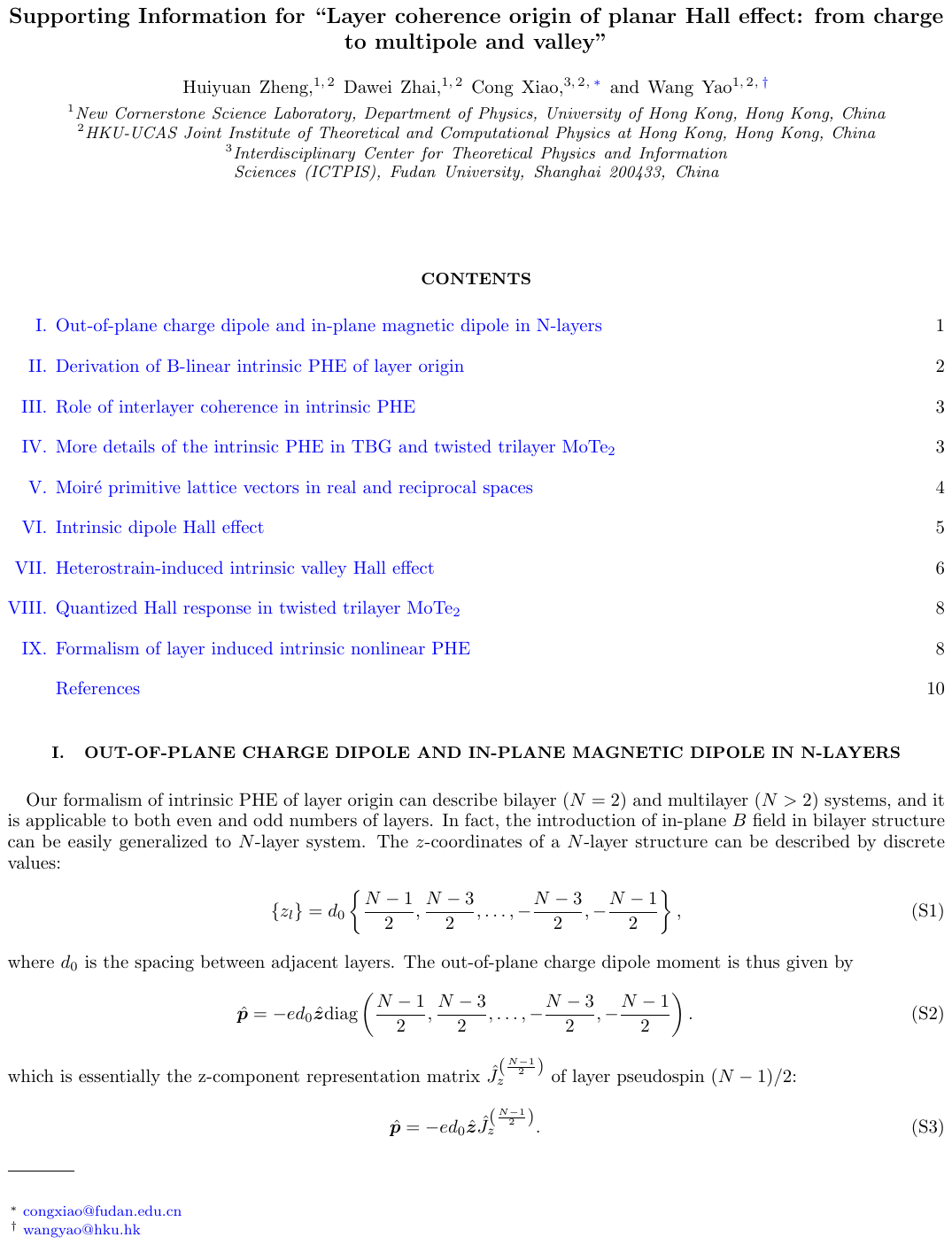}
\end{document}